# ROBUST STATE-SPACE RECONSTRUCTION OF BRAIN DYNAMICS VIA BOOTSTRAP MONTE CARLO SSA

*Sir-Lord Wiafe[1], Carter Hinsley[2], Vince D. Calhoun[1]*

[1]Tri-Institutional Center for Translational Research in Neuroimaging and Data Science (TReNDS), Georgia State University, Georgia Institute of Technology, and Emory University, Atlanta, GA 30303, USA. [2]Department of Mathematics & Statistics, Georgia State University, Atlanta, GA 30303, USA.

## ABSTRACT

Reconstructing latent state-space geometry from time series provides a powerful route to studying nonlinear dynamics across complex systems. Delay-coordinate embedding provides the theoretical basis but assumes long, noise-free recordings, which many domains violate. In many real-world domains, recordings are short, noisy, and coarsely sampled; in neuroimaging, for example, fMRI additionally contains autocorrelated background structure that can obscure oscillatory components and destabilize embeddings. We propose bootstrap Monte Carlo singular spectrum analysis (BMC-SSA), which combines Monte Carlo SSA with bootstrap stability to retain oscillatory modes that are statistically supported and reproducible across resampled data. This produces reconstructions that emphasize reliable oscillatory structure, enhancing determinism and stabilizing subsequent embeddings. Our results show that BMC-SSA improves the reliability of functional measures and uncovers differences in state-space dynamics in fMRI, offering a general framework for robust embedding of noisy, finite signals.

## 1. INTRODUCTION

Understanding the dynamics of complex systems often requires reconstructing latent state-space structure from observed signals. Delay-embedding theory, formalized by Takens [1], provides a foundation: under deterministic dynamics, sufficiently long records, and noise-free observation, delay coordinates recover a diffeomorphic image of the underlying attractor. Recent work has attempted to apply delay embeddings to functional magnetic resonance imaging (fMRI) [2]. However, resting-state fMRI strongly departs from Takens' assumptions, with short scan lengths, low-sampling-rate acquisition, hemodynamic blur, and measurement noise raising concerns that naïve embeddings may reflect acquisition artifacts rather than neural dynamics [3, 4].

A central obstacle in finite, noisy time-series analysis is that structured background fluctuations can be mistaken for dynamical organization; in fMRI, this often appears as a strong autocorrelated background structure that is only partially addressed by conventional preprocessing steps [5]. Such structure can obscure oscillatory components and bias geometric inferences in reconstructed state spaces. In general, time-series analysis shows that autocorrelated fluctuations can inflate apparent dimensionality and confound nonlinear measures if not accounted for [3, 6]. Without explicit component selection, apparent "wave-like" modes or phase-space portraits may arise from background temporal structure rather than reliable oscillatory dynamics [7]. These considerations motivate a reconstruction step that identifies oscillatory components that are statistically supported and reproducible before attempting state-space reconstruction.

We therefore introduce bootstrap Monte Carlo singular spectrum analysis (BMC-SSA) as a general front-end to delay embedding for finite, noisy signals. BMC-SSA combines Monte Carlo SSA with bootstrap stability to retain oscillatory components that are statistically supported relative to background temporal structure and reproducible across resampled data [8]. This yields reconstructions that emphasize reliable oscillatory structure rather than unstable or noise-dominated components. By emphasizing periodic and quasi-periodic content, retaining stable oscillatory modes increases determinism in the observed dynamics [9] and provides a more faithful opportunity to capture latent structure than embeddings of raw preprocessed signals alone.

We analyze this approach using resting-state fMRI by first testing whether BMC-SSA reconstructions improve within-subject reliability of standard functional connectivity measures (correlation, magnitude-squared coherence, and phase-locking value). Establishing reliability ensures that the signals provide a stable basis for subsequent embedding. We then apply Takens' delay embedding and quantify state-space structure with recurrence quantification analysis. This offers a principled and more faithful route to uncover latent dynamics from finite, noisy recordings, with fMRI serving as a challenging test case.

## 2. METHODS

### 2.1. Singular spectrum analysis (SSA)

Let a univariate time series $x(t)$, $t = 1, \dots, N$, be given. The first step in singular spectrum analysis (SSA) is to form a trajectory matrix by embedding the signal into lagged vectors:

$$\mathrm{X} = \begin{bmatrix} x(1) & \cdots & x(K) \\ \vdots & \ddots & \vdots \\ x(L) & \cdots & x(N) \end{bmatrix} \in \mathbb{R}^{L \times K}, \quad (1)$$

where $L$ is the embedding window length ($2 \leq L \leq N/2$ ) and $K = N - L + 1$. The SSA embedding window determines the time scale of resolvable components. Smaller $L$ favors short-term fluctuations, whereas larger $L$ emphasizes long-term structure. A balanced choice of $L \approx N/2$ is commonly used when there is no prior knowledge of dominant time scales, as it provides broad sensitivity to oscillatory modes, trends, and noise components [10]. We therefore adopt $L = N/2$ in all analyses.

The lag-covariance matrix is defined as

$$C = \frac{1}{K} XX^{\top}. \quad (2)$$

Performing an eigenvalue decomposition,

$$CU = U\Lambda, \quad (3)$$

yields eigenvectors $U = [u_1, \dots, u_L]$, known as empirical orthogonal functions (EOFs), and eigenvalues $\Lambda = \mathrm{diag}(\lambda_1, \dots, \lambda_L)$, which quantify the variance explained by each component.

The corresponding principal components (PCs) are obtained as projections of the original series onto the EOFs:

$$a_i(t) = \sum_{j=1}^{L} u_i(j) x(t + j - 1), \quad i = 1, \dots, L \quad (4)$$

Each pair $(u_i, a_i)$ defines an elementary component of the signal. Importantly, if the underlying data contain oscillatory structure, SSA typically produces nearly equal eigenvalue pairs $(\lambda_i, \lambda_{i+1})$. The associated EOFs are approximately in quadrature (sine–cosine phase shift), providing a data-adaptive representation of oscillatory structure.

### 2.2. Monte Carlo singular spectrum analysis (MC-SSA)

To assess the statistical relevance of SSA modes, we employ Monte Carlo SSA (MC-SSA) with an autocorrelated surrogate model for background temporal structure. The surrogate model represents the background temporal structure of the observed series $x(t)$ using a first-order autoregressive AR(1) process [11]. Surrogate time series $y(t)$ of length $N$ are generated according to

$$y(t) = \phi y(t-1) + \varepsilon(t), \quad \varepsilon(t) \sim \mathcal{N}(0, \sigma^2) \quad (5)$$

where $\phi$ is the autoregressive coefficient, and $\varepsilon(t)$ is zero-mean Gaussian innovation noise with variance $\sigma^2$. Parameters $\phi$ and $\sigma^2$ are estimated from $x(t)$ via the Yule–Walker approach, i.e., by matching the lag-1 autocorrelation and residual variance of the observed series.

From $S$ surrogate realizations, SSA is applied with the same embedding dimension $L$, yielding empirical null distributions of eigenvalues at each rank. For each rank $i$, the observed eigenvalue $\lambda_i$ is compared against the distribution of surrogate eigenvalues $\left\{\lambda_i^{(s)}\right\}$. Significance is determined by the fraction of surrogate values at rank $i$ that are greater than or equal to the observed value. Modes are retained as statistically significant when this fraction falls below a threshold of $\alpha = 0.05$.

Oscillatory SSA components typically appear as adjacent eigenvalue pairs with approximately matched variance. We therefore grouped adjacent statistically significant modes into oscillatory pairs to preserve the sine–cosine structure of oscillatory components. If either member of such a pair passed the Monte Carlo test, both were retained.

### 2.3. Bootstrap Monte Carlo SSA (BMC-SSA)

Some statistically significant SSA modes may still be unstable due to finite-sample fluctuations [11]. To ensure that retained modes are both statistically significant and reproducible, we augment MC-SSA with bootstrap stability analysis, yielding bootstrap Monte Carlo SSA (BMC-SSA).

Bootstrap resamples are generated from the observed series $x(t)$ using an overlapping moving-block bootstrap. The block length, $T$ is chosen as $T = 1/f_{min}$, so that each block spans at least one full cycle at the lowest analysis frequency. Blocks are sampled with replacement (circular wrap to avoid edge artifacts) and concatenated to produce a bootstrap replicate $x^{*(b)}(t)$ of length $N$. For each bootstrap replicate $b = 1, \dots, B$, we reapply SSA with the same embedding dimension $L$. For every mode $i$ in the original set of significant modes $\mathcal{M}$, we check whether a corresponding mode appears in the bootstrap decomposition. A mode is considered recovered when a bootstrap mode matches the original mode in both explained variance and EOF shape. In implementation, matching used a fixed relative eigenvalue tolerance of 10% and an EOF correlation threshold of 0.9.

If any bootstrap mode meets these criteria, mode $i$ is marked as recovered in replicate $b$. The stability frequency

$$\pi_i = \frac{\#\{\text{replicates recovered by mode } i\}}{B} \quad (6)$$

quantifies how reliably each significant mode reappears under resampling of the observed data. Modes with $\pi_i \geq \theta$ form the robust set $\mathcal{M}_{robust}$. Each retained SSA component is mapped back to the original time domain by diagonal averaging of its rank-1 Hankel block. The final reconstruction is obtained from the set of retained modes,

$$\hat{x}(t) = \sum_{i \in \mathcal{M}_{robust}} RC_i(t) \quad (7)$$

We set $\theta = 0.7$, consistent with stability-based resampling conventions in which recovery rates near or above 70% indicate moderate reproducibility [12], and with ICA reproducibility studies that commonly use thresholds in the 0.6–0.8 range to identify stable components [13].

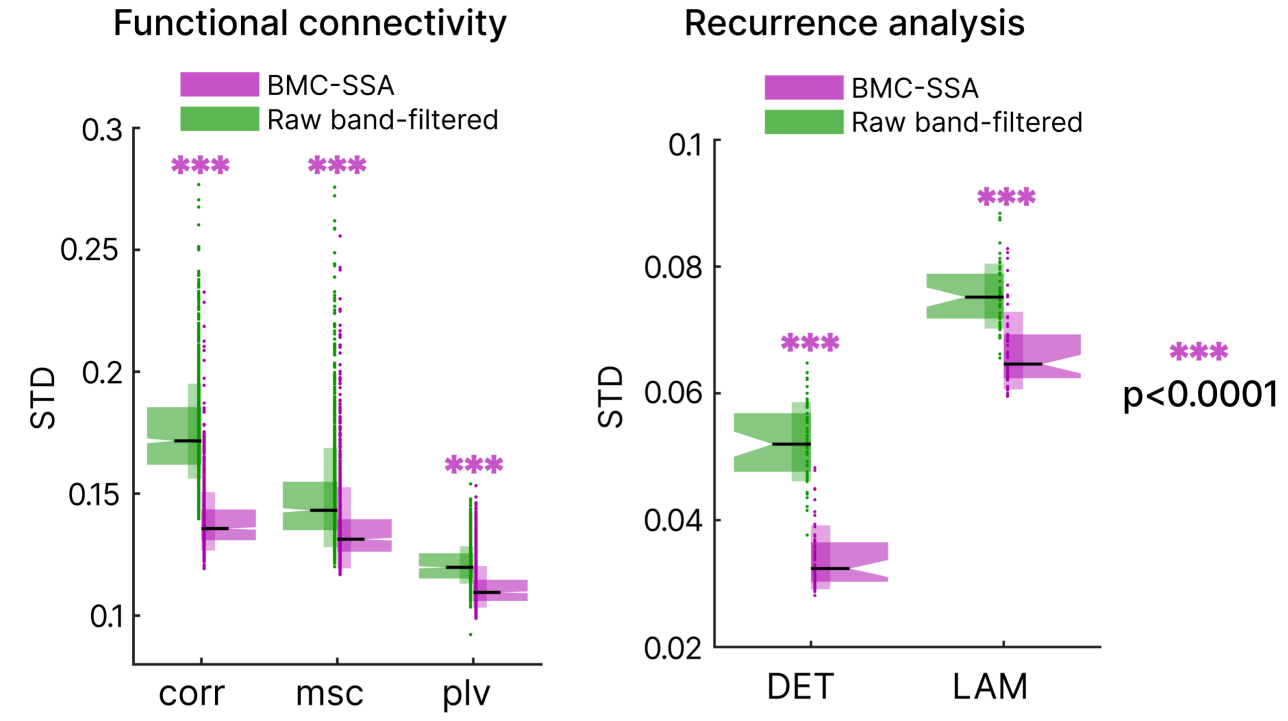

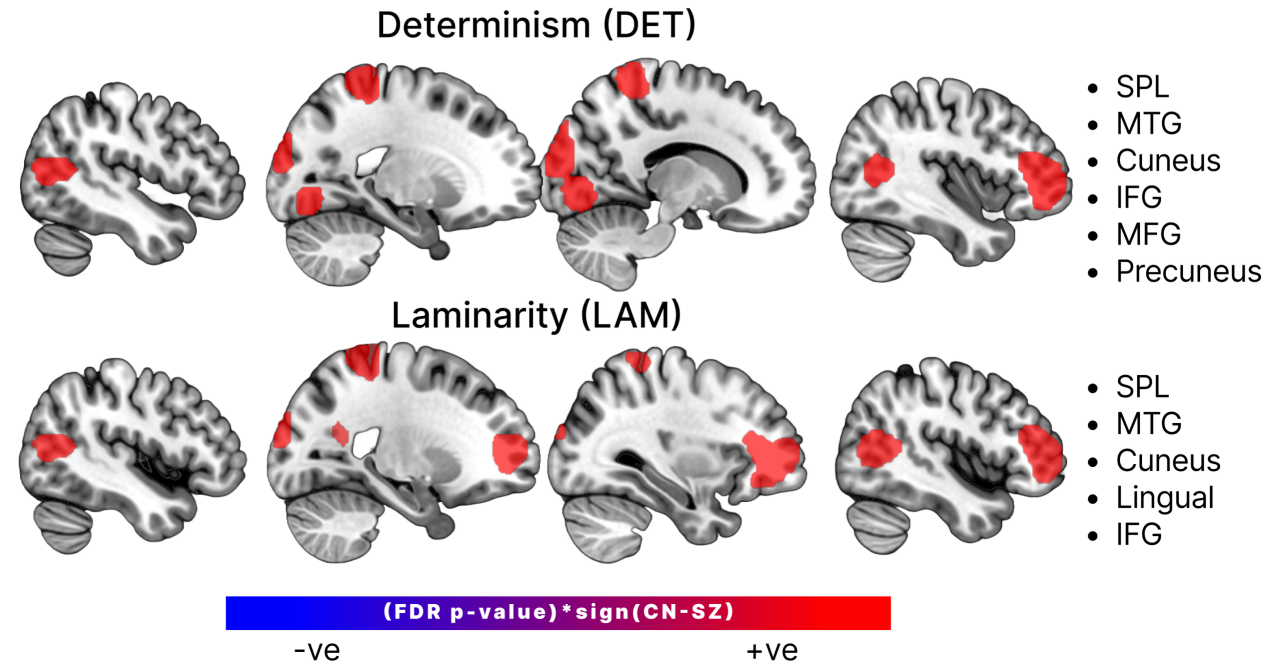

*Figure 1.* ***(a)*** *Reliability of FNC measures (corr, msc, plv) is higher (lower within-subject variability) for BMC-SSA (purple) than raw bandpass-filtered (green).* ***(b)*** *Reliability of RQA measures (DET, LAM) is also higher for BMC-SSA.* ***(c)*** *Group differences (controls > schizophrenia, FDR p<0.05) in DET and LAM, with significant regions shown on brain surfaces. These show that BMC-SSA not only stabilizes conventional connectivity metrics but also uncovers clinically relevant dynamical disruptions in schizophrenia.*

### 2.4. Dataset and Data processing

We used two complementary resting-state fMRI datasets to evaluate the proposed framework: Human Connectome Project–Development (HCP-D; N = 650 healthy subjects, TR = 0.8 s, four sessions per subject across AP/PA phase-encoding) [14] for reliability assessment, and Functional Biomedical Informatics Research Network (fBIRN; N = 311, 160 controls and 151 patients with schizophrenia, TR = 2.0 s, single session) [15] for clinical utility. Resting-state fMRI underwent standard preprocessing, after which the NeuroMark ICA pipeline extracted 53 intrinsic connectivity networks (ICNs) with the GIFT toolbox (https://trendscenter.org/software/gift/) [16]. These ICNs were despiked, detrended, bandpass filtered (0.01–0.1 Hz), and z-scored before further analyses.

**a. Representative state space (Inferior frontal gyrus)- Controls**

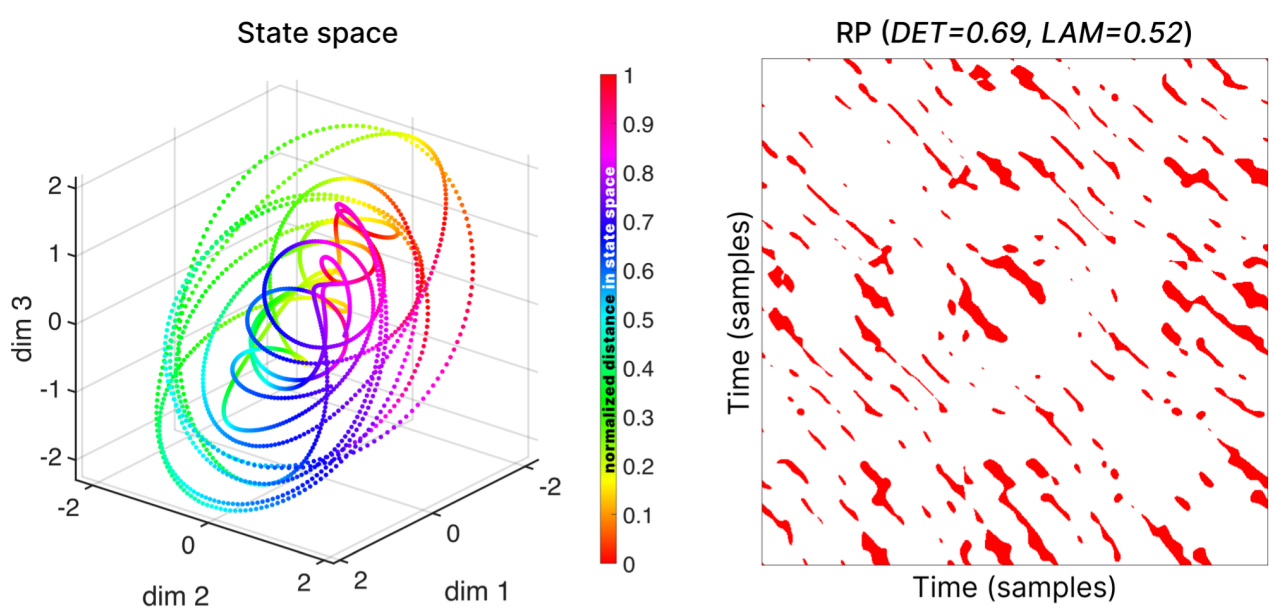

**b. Representative state space (Inferior frontal gyrus) - Schizophrenia**

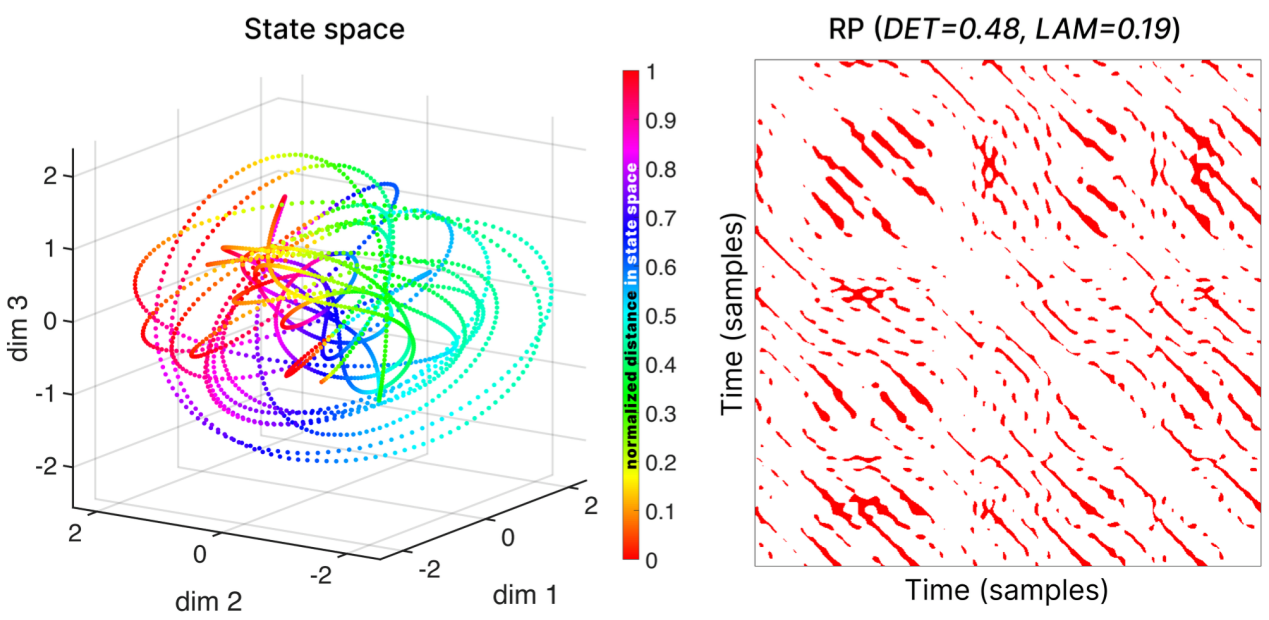

*Figure 2. After BMC-SSA, subject time series for each group were aggregated via a concatenated Hankel matrix within a region; SVD provided group-level modes from which a representative trajectory from inferior frontal gyrus was reconstructed for controls and patients. State-space embeddings and recurrence plots illustrate group-level dynamics.*

### 2.5. Functional network connectivity (FNC) reliability

To test whether the proposed pipeline yields more reliable functional estimates than conventional preprocessing, we assessed within-subject reliability across the four HCP-D resting-state fMRI sessions, comparing BMC-SSA reconstructions to raw bandpass-filtered signals. For each subject and session, ICN time courses were extracted and processed into two versions: (i) conventional (0.01–0.1 Hz bandpass) and (ii) BMC-SSA reconstruction from the robust mode set $\mathcal{M}_{robust}$. Using identical ICN pairs and time indices, we computed functional network connectivity with three measures: Pearson correlation (Fisher-z transformed), magnitude-squared coherence (MSC), and phase-locking value (PLV). For PLV, only BMC-SSA modes with dominant frequency in 0.03–0.07 Hz were used to ensure narrowband phase estimation via the Hilbert transform [17-19]. Reliability was quantified as the within-subject standard deviation across sessions for each edge and metric, then averaged across subjects. Lower dispersion indicates higher test–retest stability.

### 2.6. State-space reconstruction via Takens' delay embedding

Because BMC-SSA emphasizes statistically significant and reproducible oscillatory structure, the resulting components are empirically more amenable to stable delay-coordinate embeddings, which is desirable for preserving geometry in reconstructed state spaces under finite, noisy sampling [20]. Reconstruction of the state-space geometry requires specification of both a time delay and an embedding dimension.

The embedding delay $\tau$ was computed from the first local minimum of the average mutual information [21]:

$$I(\tau) = \int_{\hat{x}} \int_{\hat{x}(\tau)} p(\hat{x}, \hat{x}^{(\tau)}) \log \left( \frac{p(\hat{x}, \hat{x}^{(\tau)})}{p(\hat{x})p(\hat{x}^{(\tau)})} \right) d\hat{x} d\hat{x}^{(\tau)}, \quad (8)$$

where $p(\hat{x}, \hat{x}^{(\tau)})$ is the joint probability density of the reconstructed signal $\hat{x}(t)$ and its $\tau$-lagged copy, $\hat{x}^{(\tau)}$. The first local minimum of $I(\tau)$ corresponds to the time lag at which the series has lost the most redundancy while still retaining dynamical dependence; that delay is chosen as the embedding delay [21]. For each dataset, we obtained a single optimal delay by taking the mode of delays across all subjects and brain networks.

The embedding dimension $m$ was estimated using the false nearest neighbors (FNN) criterion [22]. For each candidate dimension $m$, we formed delay vectors

$$\hat{X}_m(t) = [\hat{x}(t), \hat{x}(t-\tau), \ldots, \hat{x}(t-(m-1)\tau)]^{\mathrm{T}}, \quad (9)$$

and, for every $\hat{x}_m(t)$, identified its nearest neighbor at time $t_{NN}$ in the $m$-dimensional space. We then embedded the same points in $m+1$ dimensions and tested whether the neighbor remained close. A neighbor was labeled false if either

$$\frac{|\hat{x}(t+m\tau) - \hat{x}(t_{NN}+m\tau)|}{\|\hat{X}_m(t) - \hat{X}_m(t_{NN})\|} > R_{tol} \ or \ \frac{\|\hat{X}_{m+1}(t) - \hat{X}_{m+1}(t_{NN})\|}{R_a} > A_{tol} \quad (10)$$

where $R_a$ is the attractor radius estimated as the standard deviation of $\hat{x}(t)$. We used conventional FNN tolerance settings of $R_{tol} = 15$, $A_{tol} = 2$ [22]. The fraction of false neighbors at dimension $m$ was computed as:

$$FNN(m) = \frac{\#\{\text{false neighbours at } m\}}{\#\{\text{neighbor pairs examined at } m\}} \times 100\% \quad (11)$$

For each dataset, we fixed the embedding dimension $m$ to the modal value across subjects and brain networks, ensuring comparability across individuals and regions.

Finally, the delay-coordinate state space was reconstructed with the dataset-specific optimal delay $\tau$ and the optimal dimension $m$, exactly as in Eq. (9). This embedding was then used for all subsequent analyses to ensure consistency across subjects and brain networks.

For the fBIRN application dataset, the slower sampling rate (TR = 2 s) and shorter time series reduce the sampling density of reconstructed trajectories. To mitigate this, we upsampled the BMC-SSA reconstructions by a factor of four using a polyphase FIR scheme that applies a low-pass reconstruction filter (anti-imaging/antialiasing) to the signal and compensates for the filter's group delay [23, 24]. This band-limited interpolation provides a smoother numerical representation of the reconstructed trajectory while preserving the original spectral content and avoiding spurious high frequencies, thereby maintaining the oscillatory structure emphasized by BMC-SSA and supporting more stable delay-coordinate embedding. All time-based parameters were specified in seconds and then converted to samples after resampling.

### 2.7. Recurrence Quantification Analysis (RQA)

A variety of nonlinear measures can characterize state-space dynamics, including Lyapunov exponents and correlation dimension. However, these typically require long, low-noise recordings and are difficult to estimate reliably when signals are short, noisy, or coarsely sampled [25]. Recurrence quantification analysis (RQA) is more flexible for finite, noisy data and directly summarizes geometric structure in recurrence plots [9, 26]. The recurrence plot (RP) matrix $RP_{i,j}$ is constructed from the delay-embedded trajectories by setting $RP_{i,j} = 1$ when the distance between state vectors $\hat{X}_m(i)$ and $\hat{X}_m(j)$ is below a threshold $\varepsilon$, and $RP_{i,j} = 0$ otherwise. Following standard practice, $\varepsilon$ was chosen per subject and region to impose a fixed recurrence rate of 0.05, normalizing amplitude/scale differences and facilitating comparisons [26]. When forming $RP$, we set a Theiler window of approximately twice the embedding delay ($W \approx 2\tau$) to exclude trivial near-diagonal recurrences driven by autocorrelation and the construction of delay vectors [3]. For the fMRI application, minimum line lengths were set to approximately 6 s for both diagonals and verticals so that recurrences shorter than the canonical hemodynamic response were not emphasized [27]. The threshold was computed from pairwise distances after z-scoring embedding coordinates and excluding pairs within the Theiler window.

Diagonal line structures capture determinism (DET). DET marks segments where trajectories evolve in parallel over time, i.e., portions of the signal that repeat a sequence of states [9, 26]. Let $P(\ell)$ be the histogram of diagonal line lengths $\ell$. DET is defined as:

$$DET = \frac{\sum_{\ell \geq \ell_{min}} \ell P(\ell)}{\sum_{\ell \geq 1} \ell P(\ell)}. \quad (12)$$

High DET means the system often revisits and progresses through similar ordered patterns of states, which implies sequential predictability and more ordered transitions [26]. Low DET means the system returns to similar regions but proceeds in different orders, which yields fragmented and irregular progressions [26]. In the fMRI application, higher DET suggests recurring short-lived patterns of network activity, while lower DET implies reduced predictability and more irregular movement through configurations.

Vertical line structures capture laminarity (LAM). LAM marks epochs where the system dwells in a neighborhood of state space with little forward progression, short-lived quasi-stationary phases [26]. Let $P(\nu)$ be the histogram of vertical line lengths $\nu$. LAM is defined as:

$$LAM = \frac{\sum_{\nu \geq \nu_{min}} \nu P(\nu)}{\sum_{\nu \geq 1} \nu P(\nu)} \quad (13)$$

High LAM indicates frequent quasi-stationary episodes where configurations are maintained before transitioning [26]. Low LAM points to continuous movement with little lingering, which reflects fewer stable transitions [26]. In the fMRI application, higher LAM means networks sustain locally stable states for measurable durations, while lower LAM reflects fewer such dwells and more continuous switching between configurations.

To assess reliability, we repeated the within-subject analysis used for FNC on the RQA metrics using the HCP-D dataset. DET and LAM were computed for each session and brain network pair, and reliability was summarized as the within-subject standard deviation across the four sessions, then averaged across subjects. To assess whether the reconstructed dynamical structure captured clinically relevant variation, DET and LAM were computed on the reconstructed state space from the fBIRN dataset after BMC-SSA reconstruction and upsampling. Group differences between patients and controls were tested using a general linear model with age, sex, site, and mean frame displacement as covariates. Multiple comparisons across brain regions were controlled using false discovery rate (FDR) correction.

## 3. RESULTS

Across the HCP-D dataset, BMC-SSA reconstructions yielded consistently higher within-subject reliability than raw bandpass filtered signals (Fig. 1a). For functional connectivity measures, including correlation, magnitude-squared coherence, and phase-locking value, variability across sessions was significantly reduced after BMC-SSA, reflecting improved stability of conventional pairwise connectivity estimates (Fig. 1a). A similar benefit was observed for dynamical measures derived from recurrence quantification analysis: both DET and LAM exhibited lower within-subject variability with BMC-SSA compared to raw band-filtered signals (Fig. 1a). These results show that BMC-SSA enhances the robustness of both standard connectivity measures and nonlinear dynamical descriptors when applied to short and noisy fMRI recordings.

Applying the reconstructed state-space analysis to the fBIRN dataset revealed significant differences between controls and patients with schizophrenia in both DET and LAM (Fig. 1b). Controls showed higher DET and LAM, with significant effects in the superior parietal lobule (SPL), middle temporal gyrus (MTG), cuneus, inferior and middle frontal gyri (IFG & MFG), precuneus, and lingual gyrus (Fig. 1b). These regions have been repeatedly implicated in schizophrenia, encompassing attentional control (SPL, precuneus) [28], visual integration (cuneus, lingual) [29], and fronto-temporal systems supporting higher cognition and language (IFG, MFG, MTG) [30]. In our results, higher DET in controls indicates trajectories progressing through more reproducible state sequences, whereas patients show more fragmented progressions, consistent with disrupted temporal organization reported in schizophrenia [31]. Higher LAM in controls indicates longer quasi-stationary dwell periods, aligning with the "dwells" emphasized in dFNC studies [32, 33] but here captured directly through recurrence structure. Together, these findings suggest reduced predictability and stability in schizophrenia-related brain dynamics, supporting the value of reliability-oriented reconstruction before Takens' embedding.

To provide intuition for the reconstructed dynamics, we visualized generalized state spaces for a representative region (inferior frontal gyrus). Controls showed smoother trajectories with fewer crossings, whereas patients with schizophrenia exhibited more irregular paths (Fig. 2). The recurrence plots highlight more sustained recurrences in controls, aligning with higher DET and LAM; patients with schizophrenia showed more fragmented recurrences consistent with reduced predictability and stability (Fig. 2).

## 4. CONCLUSION

We introduced BMC-SSA as a robust front-end to delay embedding, enabling more faithful reconstruction of latent dynamics by retaining statistically supported and reproducible oscillatory components. Applied to fMRI as a challenging test case, the approach improved the reliability of conventional connectivity and recurrence-based dynamical measures. It revealed clinically relevant reductions in predictability and stability of brain dynamics in schizophrenia, consistent with prior studies. A limitation is computational cost: the SSA decomposition scales as $O(N^3)$ in signal length, and BMC-SSA requires repeated surrogate and bootstrap runs, yielding overall complexity $O((S+B)N^3)$. This can be demanding for large datasets, though parallelization mitigates the burden. While demonstrated here on fMRI, the framework is broadly applicable to other domains where uncovering latent dynamics from finite, noisy signals is essential.

## 5. ACKNOWLEDGEMENT

This work was supported by the National Institutes of Health (NIH) grant (R01MH123610) and the National Science Foundation (NSF) grant #2112455.